\documentclass[aps,prb,reprint,amsmath,superscriptaddress,floatfix]{revtex4-2}

\usepackage{bm}
\usepackage{amssymb}
\usepackage{gensymb}
\usepackage{graphicx}
\usepackage[usenames]{color}

\begin{document}

\title{Magnetostatic modes and criticality in quantum-Ising materials}

\author{ R.D. McKenzie}
\email[Correspondence should be addressed to R.D.M, ]{rmck@iastate.edu}
\affiliation{Ames National Laboratory, U.S. DOE, Iowa State University, Ames, Iowa 50011, USA}
\affiliation{Department of Physics and Astronomy, Iowa State University, Ames, Iowa 50011, USA}
\date{\today}

\begin{abstract}

We analyze modes in a dipole-dipole coupled quantum-Ising material, taking into consideration shape effects present in any real magnet. We find that the soft mode governing quantum criticality in a non-ellipsoidal sample is a magnetostatic mode which nulls out the effects of dynamic-demagnetization-field fluctuations. The demagnetization field is analyzed from a microscopic perspective. Two additional modes are calculated relying solely on knowledge of the demagnetization factor of the sample, one of which has lower energy than the soft mode governing quantum criticality in the bulk of the magnet. Experimental evidence for these theoretical results is provided by analysis of electronuclear modes in LiHoF$_4$, an archetypal quantum-Ising material.

\end{abstract}

\maketitle

\section{Introduction}

The transverse-field Ising model (TFIM) exhibits a quantum-phase transition (QPT) between an ordered ferromagnetic phase and a quantum-paramagnetic phase as one tunes the applied transverse field through a quantum-critical point (QCP) \cite{DuttaBook, SuzukiBook, SachdevBook}. Assuming a needle-shaped sample for which the demagnetization field is zero, one finds a soft mode whose energy drops to zero at the QCP. A needle-shaped sample (or a uniformly-magnetized ellipsoid) is an idealization. In any real magnet, demagnetization fields and domain structure must be considered. One must ask if and how quantum criticality persists in real magnets. 

We address this here by analyzing modes of a cuboidal sample of the insulating, dipole-dipole coupled, Ising magnet LiHoF$_4$. We find that quantum criticality persists, but it is a magnetostatic (Walker) mode that softens to zero at the bulk phase transition. This inhomogeneous mode nulls out demagnetization-field fluctuations present in non-ellipsoidal samples, whereas the uniform (Kittel) mode is gapped. Remarkably, in addition to the soft mode, we are able to calculate two additional modes making use of the magnetization dynamics and the demagnetization factor of the sample.  

Research on magnetostatic modes is often focused on isotropic materials and ellipsoidal samples as the resulting equations are analytically tractable \cite{Walker1, Blocker, Bryant, Borich, Borich2}. Early research on modes in an ellipsoidal uniaxial material subject to a transverse field was carried out by Smit and Beljers \cite{Smit}. Here we consider non-ellipsoidal samples with a field transverse to the easy axis of magnetization, allowing us to analyze quantum criticality and magnetostatic modes in a realistic scenario, such as the microwave resonator experiments in Refs. \cite{LiberskySM, StampGSM}.

LiHoF$_4$ is a physical realization of the TFIM, albeit with a strong hyperfine interaction \cite{Bitko}, and with the dominant interaction between spins being dipolar. One may measure transmission of microwaves through LiHoF$_4$ to measure the soft mode \cite{LiberskySM}. Shape effects, such as demagnetization fields and domain structure, must be considered in experiments testing quantum-critical phenomena \cite{LiberskySM, Wendl}, and in the development of quantum technologies \cite{TabuchiReview, LQReview, HarderReview, Rameshti, Yan, YuanReview}.

In quantum magnonics, one considers magnetic excitations which can be measured and manipulated in, for example, microwave resonators. New technologies, including high sensitivity quantum magnetometers \cite{LQScience}, and microwave to optical transducers \cite{Hisatomi}, are rapidly developing. The primary focus of quantum magnonics has been yttrium-iron garnet (YIG), which exhibits magnetic order at room temperature \cite{Cherepanov, Serga}. LiHoF$_4$ undergoes a ferromagnetic/paramagnetic phase transition at cryogenic temperatures ($\sim 1.5K$). This material has more than three times the spin density of YIG, and at zero temperature it undergoes a QPT at a transverse field of 4.9T. This makes it an experimentally relevant testing ground for quantum effects in a many-body system, and the development of quantum technologies.

Domain structure forms to minimize the magnetostatic energy due to demagnetization fields. In a uniformly magnetized sample, the demagnetization field is due to spins at the sample surface; in a non-uniformly magnetized sample one must also consider demagnetization fields due to bulk magnetic charge. In the absence of an applied field, domains will organize themselves so that the critical behavior of the system is independent of the sample boundaries, in accord with Griffith's theorem, which states that a dipolar-coupled magnet has a well defined thermodynamic limit, independent of the sample's shape \cite{GriffithsFE}.

If an inhomogeneous ac field is applied to a uniformly magnetized material, one may excite magnetostatic modes \cite{Walker1, Walker2, Blocker}. Alternately, if a uniform ac field is applied to a non-ellipsoidal magnetic sample, the resulting demagnetization-field fluctuations may excite magnetostatic modes. We analyze these modes accounting for the exchange, dipolar, and hyperfine fields present in LiHoF$_4$. The equation of motion (EOM) of the magnetic moments, or spin operators, is used to calculate the modes, and a Walker-like equation that determines magnetostatic modes is obtained. This framework is generalized to electronuclear systems in which the soft mode is a hybridized electronuclear excitation. 

In Sec. \ref{sec:DD} we consider the dipole-dipole coupled Ising model, and make contact between the microscopic spin Hamiltonian and the macroscopic field and magnetization determined by Maxwell's equations. This elucidates the microscopic origin of the demagnetization field. In Sec. \ref{sec:MMTFIM}, we include a transverse magnetic field, and make use of the EOM for the magnetic moments to determine the modes, including magnetostatic modes, and the dynamic susceptibility of the TFIM. In addition to the soft mode, we find two modes whose energies depend on the demagnetization factor of the sample. 

The theory is generalized to include hyperfine interactions in Sec. \ref{sec:MMEN}. The framework developed in the previous sections is still valid; however, one must consider a $6 \times 6$ system of equations governing the electronuclear dynamics, and the soft mode is a hybridized electronuclear excitation. In Sec. \ref{sec:MMLiHo}, the theory is applied to LiHoF$_4$ and compared to experimental data.

\section{Dipolar sums and demagnetization fields in uniaxial magnets}
\label{sec:DD}

To begin, consider a dipole-dipole coupled Ising magnet
\begin{align}
\label{eq:Hdip}
\mathcal{H}_{dip} = -\frac{1}{2} \sum_{i \neq j} V_{ij}^{zz} J_i^z J_j^z.
\end{align}
The interaction between spins is given by $V_{ij} = J_D D_{ij}^{zz}$, where $J_D = \mu_0 (g \mu_B)^2/(4\pi)$, and the interaction is
\begin{align}
D_{ij}^{zz} = \frac{1}{r_{ij}^3} \biggr(\frac{3z_{ij}^2 }{r_{ij}^2}-1\biggr),
\end{align}
with $r_{ij} = |\boldsymbol{r}_i - \boldsymbol{r}_j|$. We allow for quantum or classical spins; however, in this preliminary section, we ignore effects due to an applied transverse field, an exchange field, and the hyperfine field. These will be included in Secs. \ref{sec:MMTFIM}, \ref{sec:MMEN}, and \ref{sec:MMLiHo}. Here we analyze the terms appearing in the dipolar field acting at site $i$ due to the rest of the spins in the material.

To make contact between the microscopic spin Hamiltonian given by Eq. (\ref{eq:Hdip}), and the macroscopic field determined by Maxwell's equations, we consider the local moments $\mu_i^z = \gamma J_i^z =  -g\mu_B J_i^z$, or the local magnetization $M_i^z = \rho_s \mu_i^z$, where $\rho_s =\mathcal{N}/V$ is the spin density. In terms of the magnetization, the Hamiltonian is
\begin{align}
\label{eq:HdipM}
\mathcal{H}_{dip} = -\frac{\mu_0}{2 \rho_s} \sum_{i} H_{dip,i}^z M_i^z,
\end{align}
where the dipolar field at site $i$ is
\begin{align}
\label{eq:Hdip2}
H_{dip,i}^z = \sum_{j (\neq i)} \Phi_{ij} M_j^z
\quad \text{with} \quad
\Phi_{ij} = \frac{D_{ij}^{zz}}{4\pi \rho_s}.
\end{align}
This dipolar field depends on both the sample shape and the magnetization of the material. To proceed, we separate the shape-independent component of the dipolar field from the rest.

First, consider a uniformly magnetized sample for which $M_j^z = M^z$. The dipolar field is then $H_{dip.i}^z = \Phi_{0,i} M^z$, where $\Phi_{0,i} = \sum_{j (\neq i)} \Phi_{ij}$. One may write $\Phi_{0,i}$ in terms of a shape-independent component and a demagnetization factor, $\Phi_{0,i} = \Phi_{\emptyset} - N_i^z$ \cite{Levy}. In an ellipsoid, the demagnetization factor is constant ($N_i^z=N^z$). More generally, the demagnetization factor will depend on the sample boundary and the choice of origin, leading to an inhomogeneous demagnetization field. Averaging over lattice sites, the magnetometric demagnetization factor of an arbitrarily shaped sample is defined by $N^z = 1/\mathcal{N} \sum_i N_i^z$ \cite{AharoniBook}, and we define $\Phi_0 = 1/\mathcal{N} \sum_i \Phi_{0,i} = \Phi_{\emptyset} - N^z$. This demagnetization factor depends on the sample shape, but it is independent of the temperature and other properties of the material.

The shape-independent component of the dipolar sum may be determined via explicit summation over a needle-shaped sample (a long thin cylinder in calculations), in which the demagnetization field is zero. One finds
\begin{align}
\label{eq:PhiEmpty}
\Phi_{\emptyset} = \sum_{j (\neq i)}^{needle} \Phi_{ij} = \frac{1}{4\pi} \biggr[\frac{4\pi}{3} + \lambda_{dip}\biggr].
\end{align}
The first term is responsible for the Lorentz local field, which comes from excluding the origin from the dipolar sum; the second term is a lattice correction \cite{AharoniBook}. If we define $\delta \Phi_{ij} = \Phi_{ij} - \Phi_{\emptyset}$ to be the interaction between a dipole at site $i$ and dipoles outside the needle, then the site-dependent demagnetization factor is $N_i^z = -\sum_j \delta \Phi_{ij}$. The dipolar field is
\begin{align}
H_{dip,i}^z = H_{\emptyset}^z + H_{d,i}^z,
\end{align}
where $H_{\emptyset}^z = \Phi_{\emptyset} M^z$ is shape independent, and the site-dependent-demagnetization field is $H_{d,i}^z = - N_i^z M^z$.  Averaging over the sample, one finds $\overline{H}_d^z \equiv 1/ \mathcal{N} \sum_i H_{d,i}^z = - N^z M^z$. As we are considering a uniformly magnetized sample, in which the bulk magnetic charge is zero ($\rho_m = -\nabla \cdot \boldsymbol{M} =0$), there will be no contribution to the demagnetization field from the bulk of the sample.

Now consider a system in which the magnetization varies throughout the sample. Assuming local ferromagnetic order in a multidomain sample, we write $M_j^z = M_{\emptyset}^z + \delta M_j^z$ where $M_{\emptyset}^z$ is the average magnetization of a small needle embedded within a particular domain. The demagnetization field acting at site $i$ is
\begin{align}
\label{eq:DemagField}
H_{d,i}^z = -N_i^z M_{\emptyset}^z + \sum_{j (\neq i)} \delta \Phi_{ij} \delta M_j^z.
\end{align}
The first term is the site-dependent demagnetization field acting on a uniformly magnetized sample of arbitrary shape; the second term contains corrections to the demagnetization field due to inhomogeneities in the magnetization. 

The average demagnetization field is
\begin{align}
\label{eq:DemagField2}
\overline{H}_d^z &= \frac{1}{\mathcal{N}} \sum_i H_{d,i}^z 
\\ \nonumber
&= - N^z M_{\emptyset}^z 
+ \frac{1}{\mathcal{N}}\sum_i \sum_{j (\neq i)} \delta \Phi_{ij} \delta M_j^z. 
\end{align}
Note that in a multidomain sample the magnetization will vary significantly across the sample and $\delta M_j^z$ need not be small. One may rewrite Eq. ({\ref{eq:DemagField2}}) as $\overline{H}_d^z = - N_{eff}^z M_{\emptyset}^z$, where
\begin{align}
\label{eq:Nfactor}
N_{eff}^z = N^z - \frac{1}{\mathcal{N}} \sum_i \sum_{j (\neq i)} \delta \Phi_{ij}
\frac{\delta M_j^z}{M_{\emptyset}^z}
\end{align}
is an effective demagnetization factor \cite{Brug, LiberskySM}. Domain structure will form to minimize the magnetostatic energy of a system, $E_m = -\mu_0/(2\rho_s)\sum_i H_{d,i}^z M_i^z$. In, for example, a cuboidal sample divided into stripe domains, one finds $H_{d,i}^z \approx 0$ \cite{Kittel1946, KooyEnz}. We expect this to hold true more generally; domain structure forms so that the demagnetization field in the bulk of the sample is small, so in the absence of an applied field $N_{eff}^z \approx 0$. As the magnetization of a sample depends on temperature and other properties of a material, such as exchange and hyperfine fields, so will $N_{eff}^z$.

In a mean-field (MF) analysis, the dipolar Hamiltonian given by Eq. (\ref{eq:HdipM}) may be written as a sum of three terms: The ground state energy, the single-ion (or MF) component, and a fluctuation term describing interactions between sites, $\mathcal{H}_{dip} = E_{gs} + \mathcal{H}_{MF} + \mathcal{H}_{fl}$. The single-ion component of the Hamiltonian is
\begin{align}
\label{eq:HdipMF}
\mathcal{H}_{MF} = -\frac{\mu_0}{\rho_s} M_{MF}^z  \biggr[ \Phi_{\emptyset} \sum_i M_i^z 
- \sum_i N_i^z M_i^z \biggr],
\end{align}
where the MF magnetization is determined self consistently from the shape-independent part of $\mathcal{H}_{MF}$,
\begin{align}
\label{eq:HdipMF0}
\mathcal{H}_{MF,\emptyset} = -\frac{\mu_0}{\rho_s} M_{MF}^z  \Phi_{\emptyset} \sum_i M_i^z. 
\end{align}
Note that in general $M_{\emptyset}^z \neq M_{MF}^z$ as fluctuations reduce the magnetization of a system from its MF value. The second term in Eq. (\ref{eq:HdipMF}) describes the MF component of the energy due to a spatially inhomogeneous demagnetization field. This energy stems from the first term in Eq. (\ref{eq:DemagField}); the second term in Eq. (\ref{eq:DemagField}) is included in $\mathcal{H}_{fl}$. The fluctuation term is given by 
\begin{align}
\mathcal{H}_{fl} = -\frac{\mu_0}{2\rho_s} \sum_{i \neq j} \Phi_{ij} \delta M_i^z \delta M_j^z,
\end{align}
where $\delta M_i^z$ now describes fluctuations about $M_{MF}^z$. The ground-state energy is $E_{gs} = \mathcal{N} \mu_0  \Phi_0 (M_{MF}^z)^2 /2$, where $\Phi_0=\Phi_{\emptyset}-N^z$ includes the demagnetization factor.

Solving for the magnetization of a sample from the microscopic Hamiltonian is not generally feasible; however, for a given magnetization pattern, one may treat moments outside of a small needle in the continuum limit. The magnetization and demagnetization field must satisfy the magnetostatic equations, $\nabla \times \boldsymbol{H}_d=0$ and $\nabla \cdot \boldsymbol{H}_d = -\nabla \cdot \boldsymbol{M}$. The surface and bulk contributions to the demagnetization field may be considered separately \cite{AharoniBook}. 

We have reviewed demagnetization fields and factors in uniaxial magnets. This is relevant to our discussion of magnetostatic modes below. In the following section we include quantum fluctuations due to an applied transverse field, and calculate the modes of the TFIM making use of the EOM of the magnetic moments, or equivalently, the Heisenberg EOM of the spin operators. We find that a uniform fluctuation induced by a uniform ac field applied along the easy axis of the material is gapped by an inhomogeneous ac demagnetization field. Nevertheless, quantum criticality persists, and the soft mode governing the phase transition is a magnetostatic mode which nulls out the effects of the ac demagnetization field. 

\section{Magnetostatic Modes in the Transverse-Field Ising Model}
\label{sec:MMTFIM}

We now consider magnetostatic modes present in the TFIM. For simplicity, we continue to neglect exchange and hyperfine fields; these will be included Sec. \ref{sec:MMEN}. The formalism developed here remains valid when these terms are included. 

Consider the spin-1/2, dipolar, TFIM:
\begin{align}
\label{eq:HTFIM}
\mathcal{H}_{TFIM} = -\frac{1}{2} \sum_{i \neq j} V_{ij}^{zz} J_i^z J_j^z - \Delta \sum_i J_i^x,
\end{align}
where $\Delta=\gamma B_x$ is the applied transverse field. The MF magnetization of the TFIM is determined self-consistently from
\begin{align}
\mathcal{H}_{MF,\emptyset} = -V_{\emptyset} \langle J^z \rangle_{MF} \sum_i  J_i^z -\Delta \sum_i J_i^x,
\end{align}
where $V_{\emptyset}$ is the interaction in a needle-shaped sample. The MF magnetization is $\boldsymbol{M}=[M^x,0,M^z] = \rho_s \gamma [\langle J^x \rangle_{MF},0,\langle J^z \rangle_{MF}]$, and the field is given by $\boldsymbol{H} = [H^x,0,H^z]$, where $\mu_0 H^x = \Delta / \gamma$, and $H^z= \Phi_{\emptyset} M^z$ (we dispense with the subscript MF on $\boldsymbol{H}$ and $\boldsymbol{M}$). 

In the presence of an applied longitudinal field $H_a^z$, the local longitudinal field is $H_{loc}^z = H_{\emptyset}^z + H_{in}^z$, where the internal field of the sample is $H_{in}^z = H_a^z + H_d^z$. To minimize the magnetostatic free energy, domain structure forms to minimize the internal field. Assuming domain walls with negligible energy and high mobility, the domain structure will adjust itself so that $H_a^z = -H_d^z$ \cite{Cooke, Mennenga}. The energy of the domain walls, and pinning potentials present in any real material, may lead to corrections to this result. In low-temperature microwave measurements of the modes in LiHoF$_4$, it was found that setting $H_d^z \approx -1.3 H_a^z$ gives reasonable agreement between experimental data and theoretical results \cite{LiberskySM}.

We now turn to the EOM determining the magnetization dynamics of the TFIM. In the absence of damping, the dynamics may be determined by the Landau-Lifshitz equation,
\begin{align}
\label{eq:LL}
d \boldsymbol{M}/dt = \gamma \mu_0 [\boldsymbol{M}(t) \times \boldsymbol{H}_{loc}(t)],
\end{align}  
where the local field is $\boldsymbol{H}_{loc} = \boldsymbol{H}_a + \boldsymbol{H}_{dip}$. We have suppressed the spatial dependence of the equation; it is to be understood that it applies locally. One may also determine the magnetization dynamics from the Heisenberg EOM of the microscopic spin operators (setting $\hbar = 1$), $d \boldsymbol{J}_i / dt = i[\mathcal{H}, \boldsymbol{J}_i] = \widehat{M} \boldsymbol{J}_i$, where the matrix $\widehat{M}$ determines the modes of the system. These two formalisms are equivalent, and will be used interchangeably.

One may expand the magnetization and the local field in fluctuations about their static values, 
\begin{align}
\boldsymbol{M}(\boldsymbol{r},t) &= \boldsymbol{M} + \delta \boldsymbol{M}(\boldsymbol{r},t) \\ \nonumber
\boldsymbol{H}_{loc}(\boldsymbol{r},t) &= \boldsymbol{H}_{loc} + \delta \boldsymbol{H}_{loc}(\boldsymbol{r},t).
\end{align}
Substituting into Eq. (\ref{eq:LL}), and neglecting interactions between fluctuations, we obtain the linearized, Landau-Lifshitz equation for the magnetic fluctuations in the random-phase approximation (RPA). The MF magnetization and the local field must satisfy $\boldsymbol{M} \times \boldsymbol{H}_{loc}=0$, and the fluctuations must satisfy (suppressing the spatial dependence)
\begin{align}
\label{eq:LLL}
\frac{d}{dt} \delta \textbf{M}(t) = 
\gamma \mu_0 \biggr[ \delta \textbf{M}(t) \times \textbf{H}_{loc}
+ \textbf{M} \times \delta \textbf{H}_{loc}(t) \biggr].
\end{align}
Equivalently, in the Heisenberg approach, one may expand the quantum spin operators in fluctuations about their mean values, and treat the fluctuations in the RPA \cite{JensenMackintosh}. 

We allow for inhomogeneous magnetization fluctuations and write $\delta \boldsymbol{M}(\boldsymbol{r},t) = \delta \overline{\boldsymbol{M}}(t) + \delta \boldsymbol{M}_{\epsilon}(\boldsymbol{r},t)$, where $\delta \overline{\boldsymbol{M}}(t)$ represents a uniform fluctuation in the material, and $\delta \boldsymbol{M}_{\epsilon}(\boldsymbol{r},t)$ describes inhomogeneous fluctuations about $\delta \overline{\boldsymbol{M}}(t)$. For now, we will consider a uniform fluctuation. We will reintroduce inhomogeneous fluctuations when we discuss magnetostatic modes.

Uniform, longitudinal, magnetic fluctuations lead to longitudinal fluctuations in the local field, 
\begin{align}
\label{eq:Hloc}
\delta H_{loc}^z(\boldsymbol{r},t) = \delta H_{\emptyset}^z(t) + \delta H_{in}^z(\boldsymbol{r},t),
\end{align}
where $\delta H_{\emptyset}^z(t) = \Phi_{\emptyset} \delta \overline{M}^z(t)$. The dynamic fluctuation in the internal field is $\delta H_{in}^z(\boldsymbol{r},t) = \delta H_a^z(t) + \delta H_d^z(\boldsymbol{r},t)$, where $\delta H_d^z(\boldsymbol{r},t)$ is a dynamic demagnetization-field fluctuation induced by the magnetic fluctuation. In non-ellipsoidal samples, this demagnetization-field fluctuation may be inhomogeneous.

Consider a single Fourier component of the applied ac field $\delta \boldsymbol{H}_a(t) = \delta \boldsymbol{H}_a e^{i\omega t}$ and the uniform magnetic fluctuation $\delta \overline{\boldsymbol{M}}(t) = \delta \overline{\boldsymbol{M}} e^{i\omega t}$. In matrix form, the equation governing the magnetization dynamics [Eq. (\ref{eq:LLL})] is given by (suppressing the spatial dependence of the local-field fluctuations)
\begin{align}
\label{eq:EOMMF}
\left[ \begin{array}{ccc}
-i\omega/\gamma \mu_0  & H_{\emptyset}^z & 0 \\ 
-H_{\emptyset}^z & -i\omega/\gamma \mu_0 & H^x  \\
0 & -H^x & -i\omega/\gamma \mu_0  \\ 
\end{array} \right]
\left[ \begin{array}{c}
\delta \overline{M}^x  \\ \delta \overline{M}^y \\ \delta \overline{M}^z \\ 
\end{array} \right] = 
\\ \nonumber
\left[ \begin{array}{ccc}
0  & M^z & 0 \\ 
-M^z & 0 & M^x  \\
0 & -M^x & 0  \\ 
\end{array} \right]
\left[ \begin{array}{c}
\delta H_{loc}^x  \\ \delta H_{loc}^y \\ \delta H_{loc}^z \\ 
\end{array} \right].
\end{align}
The determinant of the left-hand matrix determines the MF modes of the system; these are $\omega_{\parallel}^{MF}=0$ and $\omega_{\perp}^{MF} = \pm \gamma \mu_0 \sqrt{(H^x)^2+(H_{\emptyset}^z)^2}$. We find a ``longitudinal" zero mode, where by longitudinal we mean in the direction of the MF magnetization, and a ``transverse" mode describing magnetic moments precessing about their MF value. The positive and negative solutions correspond to clockwise and counter-clockwise spin precession about the MF expectation values of the magnetic moments. Assuming $\omega \neq 0$, one may invert the left-hand side of Eq. (\ref{eq:EOMMF}) to obtain the dynamic MF susceptibility of the TFIM, $\delta \overline{\boldsymbol{M}} = \widehat{\chi}_{MF}(\omega \neq 0) \delta \boldsymbol{H}_{loc}$ (see App. \ref{ap:TFIMsusc}). 

The MF susceptibility captures the response of the magnetic moments to the local field; the internal susceptibility captures the response to the internal field. The shape-independent component of the local field is given by $\delta H_{\emptyset}^z(t) = \Phi_{\emptyset} \delta \overline{M}^z(t)$. One may shift this to the left-hand side of Eq. (\ref{eq:EOMMF}) to obtain
\begin{align}
\label{eq:EOMInternal}
\left[ \begin{array}{ccc}
-i\omega/\gamma \mu_0  & H_{\emptyset}^z & 0 \\ 
-H_{\emptyset}^z & -i\omega/\gamma \mu_0 & H^x - \Phi_{\emptyset} M^x  \\
0 & -H^x & -i\omega/\gamma \mu_0  \\ 
\end{array} \right]
\left[ \begin{array}{c}
\delta \overline{M}^x  \\ \delta \overline{M}^y \\ \delta \overline{M}^z \\ 
\end{array} \right] = 
\\ \nonumber
\left[ \begin{array}{ccc}
0  & M^z & 0 \\ 
-M^z & 0 & M^x  \\
0 & -M^x & 0  \\ 
\end{array} \right]
\left[ \begin{array}{c}
\delta H_{in}^x  \\ \delta H_{in}^y \\ \delta H_{in}^z \\ 
\end{array} \right].
\end{align}
Assuming $\omega \neq 0$, one may invert the matrix in the left-hand side of the above equation to obtain the internal (shape-independent) susceptibility of the system in the RPA, $\delta \overline{\boldsymbol{M}} = \widehat{\chi}_{in}(\omega \neq 0) \delta \boldsymbol{H}_{in}$ (see App. \ref{ap:TFIMsusc}). The determinant of the left-hand matrix in Eq. (\ref{eq:EOMInternal}) determines the shape-independent, uniform, RPA modes of the material. 

Indeed, one finds the RPA modes of the system to be $\omega_{\parallel}=0$, and 
\begin{align}
\label{eq:SM}
\omega_{\perp}= \pm \gamma \mu_0 \sqrt{H^x (H^x - \Phi_{\emptyset} M^x) + (H_{\emptyset}^{z})^2}.
\end{align}
Note that if $H_{in}^z = 0$, the MF constraint, $\boldsymbol{M}_0 \times \boldsymbol{H}_{\emptyset}=0$, implies that $H^x - \Phi_{\emptyset} M^x=0$, which leads to $\omega_{\perp}= \pm \gamma \mu_0 H_{\emptyset}^z$. The internal RPA mode softens to zero at the critical point of the TFIM, where $H_{\emptyset}^z=\Phi_{\emptyset} M^z \sim (T_c-T)^{\beta=1/2}$ near the phase boundary. At $T=0$, the mode softens like $\omega_{\perp} \sim |B_x-B_x^c|^{z\nu}$ near the QCP, with MF critical exponents $z=1$ and $\nu=1/2$ \cite{SachdevBook, Yuan, Bauer}. The shape-independent MF and RPA modes of the TFIM are well known \cite{deGennes, BroutTFIM}; we have reviewed them here in order to compare MF and RPA theory with the magnetostatic-mode theory, which will be expounded below. 

The soft mode in a uniaxial magnet subject to a transverse field was calculated by Smit and Beljers in a paper that predates the TFIM and Walker's work on magnetostatic modes \cite{Smit, deGennes, Walker1}. They make use of the classical anisotropy energy and the associated anisotropy field. In a classical treatment, the anisotropy field appears in the Landau-Lifshitz equation. Our treatment is quantum mechanical, and the anisotropy is inherent in the TFIM. Care must be utilized if trying to make comparisons between results obtained using these two approaches. 
 
Note that in Monte-Carlo simulations \cite{Chakraborty, Tabei}, one may simulate a spherical sample ($N^{\mu}=1/3$ with $\mu = x, y, \text{or}\ z$), and treat the spins outside the sphere in MF theory; the susceptibility of the sphere is taken to be the ``experimental" susceptibility ($\widehat{\chi}_{exp}=\widehat{\chi}_{sph}$). The susceptibility of interest is the internal susceptibility, which corresponds to the susceptibility of a needle-shaped sample. For the longitudinal component of the susceptibility, one finds
\begin{align}
\chi_{in} = \frac{\chi_{sph}}{1-N^z \chi_{sph}} \quad \text{or} \quad 
\chi_{exp} = [\chi_{in}^{-1} + N^z]^{-1}.
\end{align}
When the internal susceptibility of a sample diverges, the experimental susceptibility is $1/N^z$. For recent work on the demagnetization factors of non-ellipsoidal samples of materials including LiHoF$_4$, see Refs. \cite{Twengstrom1,Twengstrom2}. 

Consider a longitudinal ac field with amplitude $\delta H_a^z$ and the resulting average magnetic fluctuation $\delta \overline{M}^z$. The average internal field is $\delta \overline{H}_{in}^z = \delta H_a^z + \delta \overline{H}_d^z = \delta H_a^z - N^z \delta \overline{M}^z$. Rearranging Eq. (\ref{eq:EOMInternal}) one finds 
\begin{align}
\label{eq:EOMWalker}
\left[ \begin{array}{ccc}
\frac{-i\omega}{\gamma \mu_0}  & H_{\emptyset}^z & 0 \\ 
-H_{\emptyset}^z & \frac{-i\omega}{\gamma \mu_0} & H^x - (\Phi_{\emptyset} - N^z) M^x  \\
0 & -H^x & \frac{-i\omega}{\gamma \mu_0}  \\ 
\end{array} \right]
\left[ \begin{array}{c}
\delta \overline{M}^x  \\ \delta \overline{M}^y \\ \delta \overline{M}^z \\ 
\end{array} \right] = 
\\ \nonumber
\left[ \begin{array}{ccc}
0  & M^z & 0 \\ 
-M^z & 0 & M^x  \\
0 & -M^x & 0  \\ 
\end{array} \right]
\left[ \begin{array}{c}
\delta H_a^x  \\ \delta H_a^y \\ \delta H_a^z \\ 
\end{array} \right].
\end{align}
In Eq. (\ref{eq:EOMWalker}), the mode determined by the left-hand matrix is gapped by the demagnetization-field fluctuations. One must then ask if criticality persists in a system with a non-zero demagnetization factor. We find that it does, but it is a magnetostatic mode that governs quantum criticality, rather than the uniform mode considered above.
      
The magnetostatic modes present in a magnetic system are obtained by treating the magnetization and field fluctuations in the magnetostatic approximation, viz., we impose the constraints 
\begin{align}
\label{eq:constraints}
\nabla \times \delta \textbf{H}_d = 0 \quad \text{and} \quad
\nabla \cdot \delta \textbf{H}_d = -\nabla \cdot \delta \textbf{M}.
\end{align}
We allow for inhomogeneous magnetization fluctuations. In the absence of an applied ac field, Eq. (\ref{eq:EOMInternal}) may be written $\delta \boldsymbol{M} = \chi_{in}(\omega \neq 0) \delta \boldsymbol{H}_d$. Introducing a magnetostatic potential, $\delta \boldsymbol{H}_d = - \nabla \Psi$, and making use of the internal susceptibility tensor (see App. \ref{ap:TFIMsusc}), one finds 
\begin{align}
\label{eq:Walker}
\sum_{\mu=x,y,z}(1+\chi_{in}^{\mu\mu})\frac{\partial^2 \Psi}{\partial \mu^2} 
+ 2\chi_{in}^{xz}\frac{\partial^2 \Psi}{\partial x\partial z}  
&= 0 \quad \text{inside}
\\ \nonumber
\nabla^2 \Psi &= 0 \quad \text{outside}.
\end{align}
This is an RPA generalization of the Walker equation to the TFIM. The off-diagonal components of the susceptibility tensor, $\chi^{xy} = -\chi^{yx}$ and $\chi^{zy} = -\chi^{yz}$, are antisymmetric and vanish from the Walker-mode equation (see App. \ref{ap:TFIMsusc}). The remaining off-diagonal component of the susceptibility, $\chi^{xz} = \chi^{zx}$, is symmetric, and leads to an additional term in the equation not present when an ac field is applied orthogonal to the direction of magnetization, as in Walker's seminal work \cite{Walker1}.

\begin{figure}[htp]
\centering
\includegraphics[trim={1.2cm 0cm 1cm 0.8cm},clip,width=9cm]{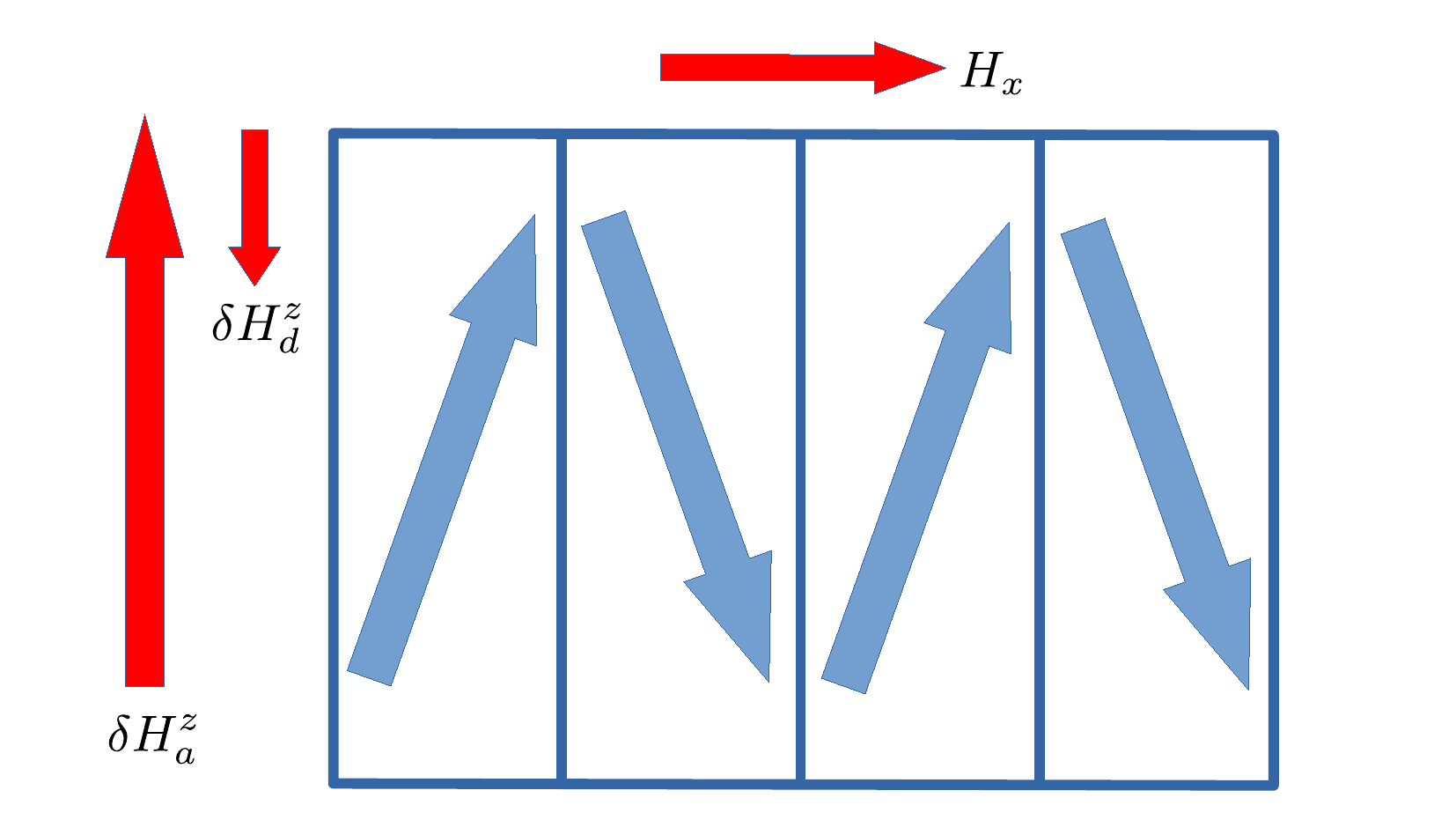}
\caption{Stripe-domain pattern in a transverse field subject to an applied longitudinal ac field and the resulting ac demagnetization field. The applied field is assumed to be uniform across the sample; however, the resulting demagnetization field need not be. }
\label{fig:DF}
\end{figure}

The allowed solutions of this Walker-like equation determine inhomogeneous magnetization fluctuation patterns, which may resonate in an appropriate applied ac field. In a uniformly magnetized ellipsoid, an inhomogeneous ac field is required to excite the magnetostatic modes \cite{Walker1, Walker2}. In a non-ellipsoidal sample, a uniform ac field may create a uniform fluctuation in the magnetization. This in turn creates an inhomogeneous ac demagnetization-field fluctuation within the sample that is responsible for the magnetostatic modes. The latter situation is relevant to the resonator experiments of interest here. In Fig. \ref{fig:DF} we show a stripe-domain pattern subject to the fields  present in the resonator experiments discussed in Refs. \cite{LiberskySM, StampGSM}.

Rather than solving Eq. (\ref{eq:Walker}), we use Eq. (\ref{eq:EOMInternal}) more directly. One approach is to make use of known solutions for $\delta \boldsymbol{H}_d$ and $\delta \boldsymbol{M}$, such as a particular domain configuration. Using the classical anisotropy energy, this approach has been utilized to study ferromagnetic resonance in barium ferrite \cite{PolderSmit, Smit, Sigal1979}. The theory has been compared to experiment with mixed success, typically at microwave frequency scales larger than the modes of interest here ($\sim 1$ GHz). It would be of interest to apply this approach to LiHoF$_4$, using the quantum formalism, but this lies beyond the scope of this work.  


Alternately, one may analyze magnetostatic modes numerically from a microscopic perspective, as in the paper by Puszkarski \textit{et al.} \cite{Puszkarski}. These researchers consider a cubic nanograin with an isotropic dipolar interaction. They find a series of magnetostatic modes including high energy bulk extended modes, and lower energy surface and bulk localized modes.

Even without detailed knowledge of $\delta \boldsymbol{H}_d$ and $\delta \boldsymbol{M}$, Eq. (\ref{eq:EOMInternal}) may be used to determine magnetostatic modes in a macroscopic system. Consider a magnetization fluctuation $\delta \boldsymbol{M}(\boldsymbol{r},t) = \delta \overline{\boldsymbol{M}}(t) + \delta \boldsymbol{M}_{\epsilon}(\boldsymbol{r},t)$. The uniform component of the magnetization fluctuation in a cuboidal sample will cause an inhomogeneous demagnetization-field fluctuation with amplitude $\delta \boldsymbol{H}_d(\boldsymbol{r})$ which gaps the soft mode. A magnetostatic mode mirroring this demagnetization field, $\delta \boldsymbol{M}_{\epsilon}(\boldsymbol{r}) = \delta \boldsymbol{H}_d(\boldsymbol{r})$, will soften to zero at the QCP. We proceed by analyzing such a mode.

First of all, one must ask if this mode satisfies the magnetostatic constraints. We find
\begin{align}
\nabla \cdot \delta \boldsymbol{M}(\boldsymbol{r}) 
= \nabla \cdot (\delta \overline{\boldsymbol{M}} + \delta \boldsymbol{M}_{\epsilon}(\boldsymbol{r})) = 0.
\end{align}
Indeed, due to the sample boundaries, the divergence of the uniform fluctuation is $\nabla \cdot \delta \overline{\boldsymbol{M}} = -\nabla \cdot \delta \boldsymbol{H}_d(\boldsymbol{r})$. By definition, 
$\nabla \cdot \delta \boldsymbol{M}_{\epsilon}(\boldsymbol{r}) = \nabla \cdot \delta \boldsymbol{H}_d(\boldsymbol{r})$. Hence, $\nabla \cdot \delta \boldsymbol{M}(\boldsymbol{r})=0$. This is consistent with a magnetostatic mode for which the demagnetization-field fluctuations are zero. This is by design. This magnetostatic mode nulls out shape effects, and softens to zero at the QCP of the TFIM.

\subsection{Magnetostatic Modes Corresponding to Static Magnetization Patterns}
\label{sec:LowerMM}

We've shown that demagnetization-field fluctuations gap the uniform mode in non-ellipsoidal samples and it is a magnetostatic mode that softens to zero at the QCP. Setting $\delta \overline{H}_d = N^z \delta \overline{M}^z$ leads to an additional mode. As will be shown in Sec. \ref{sec:MMLiHo}, there is experimental evidence for this mode, so we consider it here. We suggest this is a magnetostatic mode corresponding to static magnetization fluctuations of the material; however, this is tentative.

We assume the static magnetization of the sample varies in such a way that $|\boldsymbol{M}_i| = |\boldsymbol{M} + \delta \boldsymbol{M}_i| < |\boldsymbol{M}|$, as at, for example, a linear domain wall \cite{Bulaevskii}. Precession of this magnetization pattern will create a magnetostatic mode. At site $i$ we have $\delta \boldsymbol{M}(\boldsymbol{r}_i,t) = \delta \boldsymbol{M}_i e^{i\omega t}$, and the corresponding time dependent demagnetization-field fluctuation is, 
\begin{align}
\delta \boldsymbol{H}_d (\boldsymbol{r}_i,t) 
= \sum_{j (\neq i)} \delta \Phi_{ij} \delta \boldsymbol{M}_j e^{i\omega t}  . 
\end{align}
As $\delta \boldsymbol{M}_j$ and the resulting demagnetization field must satisfy the magnetostatic constraints, so too must $\delta \boldsymbol{M}(\boldsymbol{r},t)$ and $\delta \boldsymbol{H}_d (\boldsymbol{r},t)$; hence, $\delta \boldsymbol{M}(\boldsymbol{r},t)$ is a magnetostatic mode.

One may write the amplitude of the magnetic fluctuation as $\delta \boldsymbol{M}(\boldsymbol{r}) = -\delta \overline{\boldsymbol{M}} + \delta \boldsymbol{M}_{\epsilon}(\boldsymbol{r})$, with $\delta \overline{\boldsymbol{M}} > 0$. The average demagnetization-field fluctuation is then $\delta \overline{\boldsymbol{H}}_d (t) = N^z \delta \overline{\boldsymbol{M}}(t)$. Substitution into Eq. (\ref{eq:EOMInternal}) leads to
\begin{align}
\label{eq:EOMWalker2}
\left[ \begin{array}{ccc}
\frac{-i\omega}{\gamma \mu_0}  & H_{\emptyset}^z & 0 \\ 
-H^z & \frac{-i\omega}{\gamma \mu_0} & H^x - (\Phi_{\emptyset} + N^z) M^x  \\
0 & -H^x & \frac{-i\omega}{\gamma \mu_0}  \\ 
\end{array} \right]
\left[ \begin{array}{c}
\delta \overline{M}^x  \\ \delta \overline{M}^y \\ \delta \overline{M}^z \\ 
\end{array} \right] = 
\\ \nonumber
\left[ \begin{array}{ccc}
0  & M^z & 0 \\ 
-M^z & 0 & M^x  \\
0 & -M^x & 0  \\ 
\end{array} \right]
\left[ \begin{array}{c}
\delta H_a^x  \\ \delta H_a^y \\ \delta H_a^z \\ 
\end{array} \right].
\end{align}
If such a mode exists, it has lower energy than the soft mode governing bulk criticality, and it softens to zero above and below $B_x^c$. The nature of this mode is unclear; however, we expect it to lead to local regions of magnetic order/disorder away from the phase transition in which the bulk of the material becomes magnetized.


\section{Electronuclear Magnetostatic Modes}
\label{sec:MMEN}

We now generalize the results of Sec. \ref{sec:MMTFIM} to include a hyperfine coupling between each electronic spin and its nucleus. Seminal research on electronuclear modes was carried out by de Gennes \textit{et al.}, and electronuclear magnetostatic modes were first analyzed by Blocker \cite{deGennesNMR, Blocker} (see App. \ref{ap:FP}). In their work, the hyperfine interaction was assumed to be isotropic, and the applied ac field was assumed to be transverse to the direction of magnetization of the sample. Here, we allow for an anisotropic hyperfine interaction, and consider an ac field along the easy axis of the material, as is required to observe the soft mode in the TFIM \cite{MckenzieStamp, LiberskySM}.

Nuclear magnetic moments are small, and do not contribute significantly to the magnetization of a material; however, a strong hyperfine coupling dramatically impacts the magnetization dynamics. With a strong hyperfine coupling, there will be hybridized electronuclear modes, rather than the strictly electronic modes considered in Sec. \ref{sec:MMTFIM}. It is a low energy electronuclear mode that softens to zero at the QCP, and governs quantum criticality. This mode has been the subject of recent theoretical work, and has been measured experimentally \cite{MckenzieStamp, LiberskySM}.

Consider $\mathcal{H}=\mathcal{H}_{TFIM} + \mathcal{H}_{hyp}$, where $\mathcal{H}_{TFIM}$ is given in Eq. (\ref{eq:HTFIM}), and the hyperfine component of the Hamiltonian is
\begin{align}
\label{eq:hyp}
\mathcal{H}_{hyp} = \Delta_n &\sum_{i} I_{i}^{x}
+ A_{z} \sum_{i} I_{i}^{z}J_{i}^{z} 
\\ \nonumber
&+ A_{\perp} \sum_{i} (I_{i}^{x}J_{i}^{x} + I_{i}^{y}J_{i}^{y}).
\end{align}
In materials such as LiHoF$_4$, the effective transverse field acting directly on the nuclear spins may be substantial, even though the applied transverse field couples weakly to the nuclei \cite{MckenzieStamp}. As in Sec. \ref{sec:MMTFIM}, one may treat the interaction between electronic spins in the MF approximation, and, considering a needle-shaped sample, self-consistently determine both the electronic and nuclear spin polarizations.

The EOM for the electronic spin operators may be written
\begin{align}
\frac{d}{dt} \boldsymbol{J}_i  
&= i [\mathcal{H}_{TFIM},\boldsymbol{J}_i] + i [\mathcal{H}_{hyp},\boldsymbol{J}_i]
\\ \nonumber
&= \widehat{M}_{TFIM}\ \boldsymbol{J}_i  + \widehat{M}_{hyp} \ \boldsymbol{J}_i,
\end{align}
where
\begin{align}
\widehat{M}_{TFIM} = \left[ \begin{array}{ccc}
0   &  \sum_j V_{ij} J_j^z  & 0          \\
- \sum_j V_{ij} J_j^z & 0    & \Delta    \\
0  &  -\Delta  &  0    \\
\end{array} \right],
\end{align}
and the hyperfine component of the equation governing the electronic spin dynamics is
\begin{align}
\widehat{M}_{hyp}  = \left[ \begin{array}{ccc}
0   &  -A_z I_i^z  & A_{\perp} I_i^y          \\
A_z I_i^z & 0    & -A_{\perp} I_i^x    \\
-A_{\perp} I_i^y  &  A_{\perp} I_i^x  &  0    \\
\end{array} \right].
\end{align}
The EOM for the nuclear spins is
\begin{align}
\frac{d}{dt} \boldsymbol{I}_i  
= i [\mathcal{H}_{hyp},\boldsymbol{I}_i] = \widehat{M}_{hyp}^{I} \ \boldsymbol{I}_i,
\end{align}
with
\begin{align}
\widehat{M}_{hyp}^{I} = \left[ \begin{array}{ccc}
0   &  -A_z J_i^z  & A_{\perp} J_i^y          \\
A_z J_i^z & 0    & -\Delta_n -A_{\perp} J_i^x   \\
-A_{\perp} J_i^y  &  \Delta_n + A_{\perp} J_i^x  &  0    \\
\end{array} \right].
\end{align}
These coupled equations determine the electronuclear dynamics of the full Hamiltonian. To proceed, we decouple the interactions between spins in the RPA.

To capture the electronuclear dynamics of the coupled system we consider the combined operator $\boldsymbol{X}_i = (\boldsymbol{J}_i, \boldsymbol{I}_i)$, and write the pair of $3 \times 3$ matrices governing the dynamics as a single $6 \times 6$ matrix. In the RPA, we consider fluctuations of the electronic and nuclear spins about fixed values, $\boldsymbol{J}_i(t) = \langle \boldsymbol{J} \rangle + \delta \boldsymbol{J}_i(t)$ and $\boldsymbol{I}_i(t) = \langle \boldsymbol{I} \rangle + \delta \boldsymbol{I}_i(t)$, and drop interactions between fluctuations. 

From the MF component of the EOM one finds $\boldsymbol{H}_e \times \langle \boldsymbol{J} \rangle = 0$. The components of the electronic MF are
\begin{align}
\label{eq:ElecMF}
\gamma_e \mu_0 H_e^z &=  V_{\emptyset} \langle J^z \rangle - A_z \langle I^z \rangle 
\\ \nonumber
\gamma_e \mu_0 H_e^y &=  -A_{\perp} \langle I^y \rangle
\\ \nonumber
\gamma_e \mu_0 H_e^x &= \Delta - A_{\perp} \langle I^x \rangle.
\end{align}
For the case at hand, $\langle I^y \rangle = 0$. The hyperfine interaction shifts the longitudinal and transverse fields acting on the electronic spins. For the nuclear spins, one finds $\boldsymbol{H}_n \times \langle \boldsymbol{I} \rangle = 0$, where the MF acting on the nuclear spins is defined by
\begin{align}
\label{eq:NucMF}
\gamma_n \mu_0 H_n^z &= -A_z \langle J^z \rangle \\ \nonumber
\gamma_n \mu_0 H_n^y &= -A_{\perp} \langle J^y \rangle \\ \nonumber
\gamma_n \mu_0 H_n^x &= -\Delta_n - A_{\perp} \langle J^x \rangle.
\end{align}
These coupled equations determine the MF electronic and nuclear spin polarizations. With $\langle J^y \rangle = \langle I^y \rangle = 0$ one finds
\begin{align}
\label{eq:MFeq}
\langle J_x \rangle H_e^z = \langle J_z \rangle H_e^x 
\quad \text{and} \quad
\langle I_x \rangle H_n^z = \langle I_z \rangle H_n^x.
\end{align}
One may solve these equations to determine the MF electronic and nuclear spin polarizations; however, there is a discrepancy between the result obtained from Eq. (\ref{eq:MFeq}) and the values determined self-consistently from the MF Hamiltonian. In what follows, we use the result obtained from the MF Hamiltonian, as this approach avoids the electronuclear decoupling necessary in the EOM approach.

In the RPA, the EOM governing the electronuclear spin fluctuations is given by
\begin{align}
\label{eq:RPA}
\frac{d}{dt} \delta \boldsymbol{X}_i = \widehat{M}_{RPA}\ \delta \boldsymbol{X}_i 
= \left[ \begin{array}{cc}
\widehat{M}_e   &  \widehat{M}_{en}          \\
\widehat{M}_{ne} & \widehat{M}_n            \\
\end{array} \right]
\left[ \begin{array}{c}
\delta \boldsymbol{J}_i \\
\delta \boldsymbol{I}_i \\ 
\end{array} \right].
\end{align}
Setting $\langle J^y \rangle = \langle I^y \rangle = 0$, and making use of Eqs. (\ref{eq:ElecMF}) and (\ref{eq:NucMF}), one finds the matrix governing the electronic sector of the time evolution to be
\begin{align}
\label{eq:Me}
\widehat{M}_e = \left[ \begin{array}{ccc}
0   &  H_e^z  & 0  \\
- H_e^z & 0  & H_e^x - V_{\emptyset} \langle J^x \rangle     \\
0  &  -H_e^x  &  0    \\
\end{array} \right].
\end{align}
This is equivalent to the result one obtains for the TFIM, except that the MF acting on the electronic spins contains a hyperfine correction. The electronic spin fluctuations are coupled to the nuclear spin fluctuations through 
\begin{align}
\widehat{M}_{en} = \left[ \begin{array}{ccc}
0   &  A_{\perp} \langle J^z \rangle  & 0          \\
-A_{\perp} \langle J^z \rangle & 0    & A_z \langle J^x \rangle   \\
0  &  -A_{\perp} \langle J^x \rangle  &  0    \\
\end{array} \right].
\end{align}
This  coupling hybridizes the electronic and nuclear spin fluctuations leading to electronuclear modes. Comparing this result with the dynamical equations in Section \ref{sec:MMTFIM}, it is apparent that the nuclear spins drive the electronic spins, with the nuclear-drive field being $\delta h_n^z(t) = (A_{\perp} \delta I^x(t), A_{\perp} \delta I^y(t), A_z \delta  I^z(t))$.

The matrix governing the nuclear sector of the EOM is given by
\begin{align}
\widehat{M}_{n} = \left[ \begin{array}{ccc}
0   &  H_n^z & 0          \\
-H_n^z & 0    & H_n^x  \\
0  &  -H_n^x  &  0    \\
\end{array} \right].
\end{align}
This leads to nuclear spin precession about a MF created by the polarized electronic spins. The nuclear spins couple to, or are driven by, the electronic spins via
\begin{align}
\widehat{M}_{ne} = \left[ \begin{array}{ccc}
0   &  A_{\perp} \langle I^z \rangle  & 0          \\
-A_{\perp} \langle I^z \rangle & 0    & A_z \langle I^x \rangle   \\
0  &  -A_{\perp} \langle I^x \rangle  &  0    \\
\end{array} \right].
\end{align}
One may diagonalize the $6 \times 6$ EOM governing the electronic and nuclear spin fluctuations [Eq. (\ref{eq:RPA})] to obtain the electronuclear modes of the system.

Our analysis has been carried out for a needle-shaped sample. The resulting modes are the shape independent, uniform, RPA modes of the material, analagous to the modes determined by Eq. (\ref{eq:EOMInternal}). One may account for time dependent demagnetization-field fluctuations and magnetostatic modes in the same manner as in Sec. \ref{sec:MMTFIM}. This leads to a shift in $V_{\emptyset}$:
\begin{align}
V_{\emptyset} = \rho_s J_D \frac{D_{\emptyset}^{zz}}{\rho_s}
= \rho_s J_D \biggr(\frac{4\pi}{3} + \lambda_{dip}\biggr) 
\\ \nonumber
\rightarrow V_0 = \rho_s J_D 
\biggr(\frac{4\pi}{3} + \lambda_{dip} \pm 4\pi N^z\biggr).
\end{align}
In addition to the (magnetostatic) mode which softens to zero at the QCP, there is a low-energy electronuclear mode which is gapped at the critical point by the demagnetization-field fluctuations, and a magnetostatic mode with lower energy than the soft mode.

\section{Magnetostatic Modes in L$\text{i}$H$\text{o}$F$_4$}
\label{sec:MMLiHo}

We now consider electronuclear modes in LiHoF$_4$ \cite{Schechter,Schechter2, MckenzieStamp, LiberskySM}. If a transverse field is applied to a uniformly-magnetized needle, one finds an electronuclear soft mode whose energy drops to zero at the QCP. This marks a ferromagnetic to quantum-paramagnetic phase transition. The upper-critical dimension of a dipolar-Ising ferromagnet is $d^*=3$; this leads to MF critical exponents with logarithmic corrections \cite{Nikkel}.

The millimeter scale samples of LiHoF$_4$ considered here are well above the critical size required for a multidomain sample \cite{KittelDomainRMP}. The critical plate thickness necessary for the formation of domains in the dipolar-Ising magnet LiTbF$_4$ has been estimated to be $\sim 10^{-6}$ m \cite{Barker}. Optical measurements \cite{Battison, Meyer, Pommier}, and scanning-hall-probe microscopy \cite{Jorba}, show micron-scale domains in LiHoF$_4$. At low temperatures, thick plates show branched-stripe-domain structure, giving way to a bubble phase in an applied longitudinal field. Monte-Carlo simulations support the formation of stripe domains \cite{Biltmo}.  We note that in LiTbF$_4$ there is a critical speeding up of the relaxational dynamics of linear domain walls as one approaches $T_c$ \cite{Kotzler}.  

Our analysis is carried out for a needle embedded within a particular domain. The local demagnetization field [Eq. \ref{eq:DemagField}] includes contributions from domain walls, neighboring domains, as well as spins at the sample surface. Hence, our analysis is assumed to allow for domain-wall fluctuations.


At low temperatures, one may neglect all but the two lowest electronic eigenstates of LiHoF$_4$ and write down a truncated Hamiltonian (dropping a constant contribution to the ground state energy) \cite{Chakraborty, Tabei, MckenzieStamp}
\begin{align}
\label{eq:liho}
\mathcal{H}_{LiHoF_4} = - \frac{\Delta}{2} \sum_{i}\sigma_{i}^{x} 
- \frac{1}{2}  \sum_{i \neq j} V_{ij} \sigma_{i}^{z}\sigma_{j}^{z} + \mathcal{H}_{hyp},
\end{align}
where the interaction includes a dipolar component and antiferromagnetic superexchange
\begin{align}
V_{ij} = J_{D}C_{zz}^2D_{ij}^{zz} - J_{nn} C_{zz}^2 \delta_{\langle ij \rangle}.
\end{align}
Mixing of the crystal-field eigenstates by the transverse field is incorporated by the truncation procedure. The Pauli operators appearing in Eq. (\ref{eq:liho}) are related to the electronic spin operators of LiHoF$_4$ by $J^{\mu} = C_{\mu} + \sum_{\nu=x,y,z} C_{\mu\nu}\sigma^{\nu}$. The effective transverse field splitting the two lowest electronic eigenstates, $\Delta$, and the truncation parameters, $C_{\mu}$ and $C_{\mu \nu}$, are all functions of the applied transverse field.

This truncation procedure captures the low-energy modes of LiHoF$_4$ well at the MF/RPA level; however, interactions between fluctuations will shift the mode energies \cite{Ronnow}. In addition, Monte-Carlo simulations have shown that inclusion of off-diagonal-dipolar (ODD) interactions, which are negligible at the MF/RPA level due to the lattice symmetry, lead to better quantitative agreement between experimental and theoretical determinations of the system's phase diagram \cite{Dollberg, Dollberg2}. We neglect these effects here.

In LiHoF$_4$, the Land{\'e} g-factor is $g=5/4$, and there are four spins per unit cell with volume $V_{cell}=a^2c$, where the transverse lattice spacing is $a=5.175${\AA}, and the longitudinal lattice spacing is $c=10.75${\AA}. The corresponding spin density is $\rho_s = 1.39 \times 10^{28} m^{-3}$, and the dipolar energy scale is $\rho_s J_D  = 13.52mK$. We assume an antiferromagnetic superexchange interaction of $J_{nn} = 1.16mK$ between each holmium ion and its four nearest neighbours \cite{Ronnow}. 

The hyperfine coupling in rare-earth elements is primarily due to dipole-dipole interactions and spin-orbit coupling, with the coupling in holmium being the strongest of the group \cite{BleaneyHyperfine}. The hyperfine component of the LiHoF$_4$ Hamiltonian is given by $\mathcal{H}_{hyp} = A \sum_i \boldsymbol{J}_i \cdot \boldsymbol{I}_i$, where each electronic spin is coupled to an $I=7/2$ nuclear spin with strength $A=39.8$ mK \cite{Magarino2}. Substituting in the truncated electronic spin operators, one finds an anisotropic hyperfine interaction similar to Eq. (\ref{eq:hyp}), with $A_z \approx 200$ mK. This splits the electronic ground state into GHz scale electronuclear modes ($1$ GHz $\sim$ $50$ mK), the lowest of which is of interest to us here. Note that this hyperfine splitting is orders of magnitude less than typical crystal-field splittings of electronic rare-earth eigenstates. In addition to the terms appearing in Eq. (\ref{eq:hyp}), the hyperfine component of the truncated LiHoF$_4$ Hamiltonian contains a weak field acting on the nuclear spins in the $y$ direction, and a term with the form $A_{++} \sum_i \sigma_i^+ I_i^+ + h.c.$, where $\sigma_i^{\pm}$ and $I_i^{\pm}$ are the usual spin raising and lowering operators. These terms are easily incorporated in a numerical calculation of the RPA modes.

To calculate the RPA modes of the system, rather than using the EOM discussed above, we make use of the imaginary time ordered, longitudinal, Green function, $G_{ij}^{zz}(\tau) = -\langle T_{\tau} J_i^z(\tau) J_j^z(0) \rangle$ (see App. \ref{ap:TFIMsusc}). This avoids a decoupling of the electronic and nuclear spins necessary in the EOM approach. Fourier transforming to momentum and Matsubara frequency space, the Green function is related to the dynamic susceptibility of the electronic spins by $\chi_{\boldsymbol{k}}^{zz}(\omega) = -\beta G_{\boldsymbol{k}}^{zz}(i\omega_r \rightarrow \omega + i0^+)$. In the RPA, the dynamic susceptibility of the electronic spins is given by $\chi_{\boldsymbol{k}}^{zz}(\omega) = \chi_{\emptyset}^{zz}(\omega)/[1-V_{\boldsymbol{k}}\chi_{\emptyset}^{zz}(\omega)]$ where $\chi_{\emptyset}^{zz}(\omega)$ is the shape-independent, single-ion (MF) susceptibility of the system. The RPA modes of the system follow from the poles of this function. At zero wavevector, the interaction is given by 
\begin{align}
V_0 = \rho_s J_D C_{zz}^2
\biggr(\frac{4\pi}{3} + \lambda_{dip} + 4\pi \lambda_{ex} \pm 4\pi N^z \biggr).
\end{align}
The exchange contribution to the interaction is $\lambda_{ex} = -4J_{nn}/(4\pi \rho_s J_D) = -2.73 \times 10^{-2}$ and the lattice correction is $\lambda_{dip} = 1.66$. With $N^z=0$ one obtains the soft mode which governs bulk quantum criticality; with $-4\pi N^z$ one obtains a uniform mode which is gapped by dynamic-demagnetization-field fluctuations, and $+4\pi N^z$ leads to the magnetostatic mode with lower energy than the soft mode. The LiHoF$_4$ sample under consideration has dimensions $(L_x \times L_y \times L_z) = (1.8mm \times 2.5mm \times 2mm)$, and a demagnetization factor of $N^z=0.344$ \cite{Aharoni}.

\begin{figure}[htp]
\centering
\includegraphics[width=9cm]{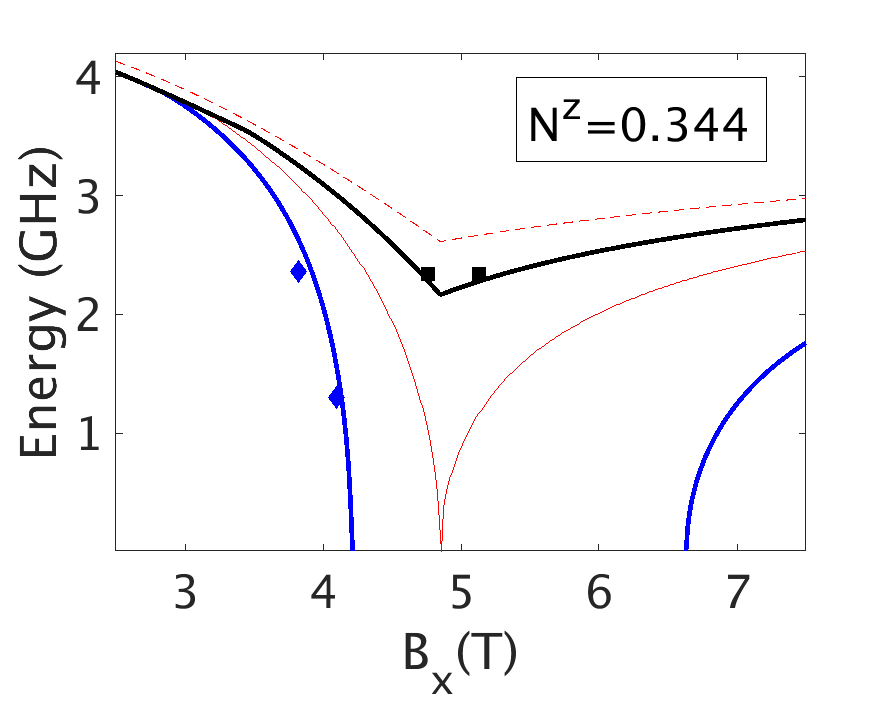}
\caption{Low energy electronuclear modes of LiHoF$_4$ at $T=50$ mK as a function of transverse field. The blue diamonds and black squares correspond to features in experimental data for a sample with $N^z=0.344$ (see Fig. \ref{fig:MM2}). The dashed red line shows the lowest single-ion excitation calculated using MF theory. Below this energy, the modes are collective excitations of the spins. The red line is the soft mode which governs bulk quantum criticality. The black line is a uniform mode gapped by demagnetization-field fluctuations. The blue line is the magnetostatic mode discussed in Sec. \ref{sec:LowerMM}.}
\label{fig:MM}
\end{figure}

\begin{figure}[htp]
\centering
\includegraphics[width=9cm,trim={0cm 0cm 5.5cm 0cm},clip]{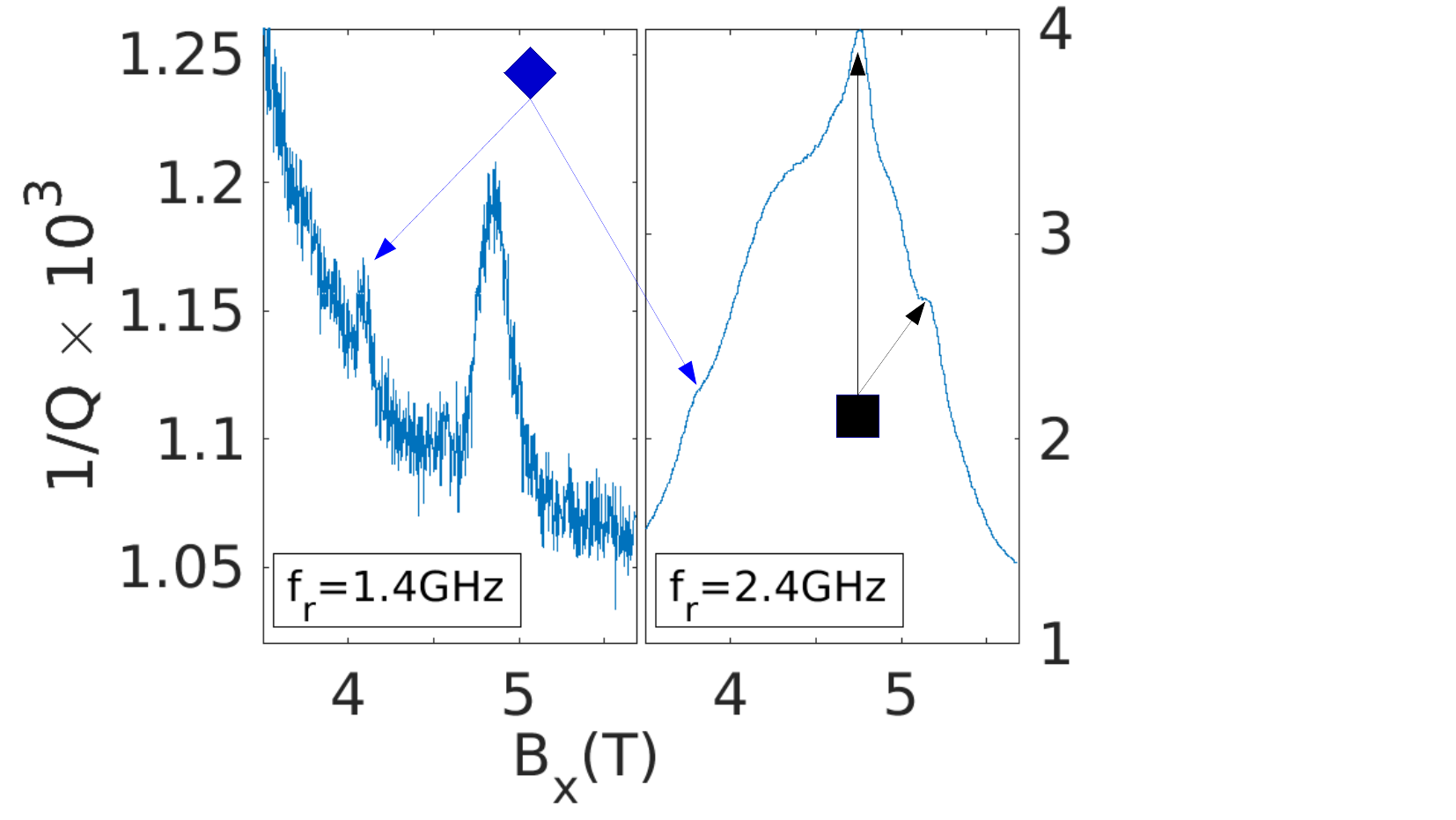}
\caption{Linewidth ($1/Q$) of the magnon-polariton mode of LiHoF$_4$ at $T=50$ mK in a $f_r = 1.4GHz$ and $f_r=2.4GHz$ resonator, plotted as a function of transverse field. The black squares indicate a uniform (Kittel) mode which is gapped by demagnetization-field fluctuations. The blue diamonds correspond to a magnetostatic mode having lower energy than the soft mode which governs bulk quantum criticality (see Sec. \ref{sec:LowerMM}). These data points are plotted along with the low-energy-electronuclear modes of a sample of LiHoF$_4$ with $N^z=0.344$ in Fig. \ref{fig:MM}.}
\label{fig:MM2}
\end{figure}

In Fig. \ref{fig:MM}, we plot low frequency modes of LiHoF$_4$ calculated at the experimentally relevant temperature of $T=50$ mK. The MF/RPA theory overestimates the critical transverse field by $0.42$ Tesla. Interactions between fluctuations neglected in the RPA will lower the result for $B_x^c$. In the following, to compare the calculated data with experimental results, we simply downshift the transverse field, $B_x \rightarrow B_x-0.42T$. We emphasize the calculated modes are determined using the theoretically determined value of the demagnetization factor. These modes are in good agreement with features of the measured absorption spectra shown in Fig. \ref{fig:MM2}. The dashed red line in Fig. \ref{fig:MM} shows the lowest single-ion excitation calculated using MF theory. In the RPA, this mode softens to zero at the critical point of the system (the solid-red line). In a needle-shaped sample ($N^z=0$), the soft mode is a uniform mode, whereas in non-ellipsoidal samples, the soft mode is a magnetostatic mode. With $N^z \neq 0$, the uniform mode is gapped by demagnetization-field fluctuations, leading to the upper mode shown in black. The lower mode shown in blue is a magnetostatic mode stemming from variations in the system's magnetization.

The blue diamonds and black squares shown in Fig. \ref{fig:MM} correspond to features seen in the inverse quality factor of a polariton mode present when LiHoF$_4$ is placed in a microwave resonator. The magnon-polariton propagator, $D_{mp}(\tau) = \langle T_\tau [a^{\dagger}(\tau)+a(\tau)](a^{\dagger}+a) \rangle$, determines propagation of photons through the resonator \cite{Mckenzie}. In Matsubara frequency space ($z=i\omega_n=i2\pi n/\beta$), one finds that
\begin{align}
D_{mp}(z) = -\frac{2\omega_r}{\beta}\biggr[\frac{1}{z^2-\omega_r^2+g_c^2 \chi(z)}\biggr].
\end{align}
The (longitudinal) dynamic susceptibility of the spins is given by $\chi(z) = \chi_{\boldsymbol{k}=0}^{zz}(z)$, and the coupling, $g_c = \eta \omega_r \sqrt{4\pi \rho_s J_D} $, is proportional to the filling factor of the resonator, $\eta$, and its frequency. The poles of this function determine the magnon-polariton modes which transmit light through the resonator.

For a resonator with frequency $\omega_r = 2\pi f_r$, the measured magnon-polariton mode follows from $\omega_{mp} = \sqrt{\omega_r^2-g_c^2 \chi'(\omega_{mp})}$. The real component of the dynamic susceptibility reduces the resonant frequency of the resonator. Setting $z=i\omega_n \rightarrow \omega+i\Gamma_r/2$, where $\Gamma_r$ is the bare linewidth of the resonator mode, one finds the linewidth of the polariton mode to be $\Gamma_{mp} = \Gamma_r+g_c^2 \chi''(\omega_{mp})/\omega_{mp}$. The absorptive component of the dynamic susceptibility increases the linewidth of the magnon-polariton mode, and hence, its inverse quality factor $1/Q= \Gamma_{mp}/\omega_{mp}$. The inverse quality factors for resonators with $f_r = 1.4GHz$ and $f_r = 2.4GHz$ are shown in Fig. \ref{fig:MM2} \cite{LiberskyUP}.

In the $f_r=1.4GHz$ data set there is a clear resonance at the expected frequency of the lower magnetostatic mode, viz., at the transverse field value where the lower magnetostatic mode cuts through the magnon-polariton mode. The more prominent peak to its right corresponds to absorption by the soft mode and at the critical point of the system. In the $f_r = 2.4GHz$ data set there is overlapping absorption due to different modes, confusing the identification of the resonances. The rightmost shoulder in the $2.4GHz$ data set is in good agreement with the gapped uniform mode in the paramagnetic phase of the system, and the dominant peak agrees well with this mode in the ferromagnetic phase. The broad shoulder left of the dominant peak in the $2.4GHz$ data set corresponds to the soft mode, and the lower shoulder, indicated by a blue diamond, is in good agreement with the lower magnetostatic mode calculated by the theory.

The experimental data sets shown were chosen because they provide the best evidence for the modes calculated in this paper. In other data sets, for resonators at other frequencies, the evidence for the modes shown in blue in Fig. \ref{fig:MM} is less compelling. As previously noted, identification of the modes is difficult when there are multiple overlapping resonances. It should also be noted that experimental signatures of the magnetostatic modes weaken as the temperature is increased \cite{StampGSM}. The magnitude of the resonances will depend on the spectral weights and the coherences of the modes, which we have not considered here. 

In the experimental data \cite{StampGSM}, we see resonances in the absorption spectrum corresponding to both single-ion (MF) and collective-mode excitations. In Ref. \cite{Mckenzie}, it was shown that assuming spectral weight is transferred from the collective (RPA) mode spectrum to the single-ion (MF) spectrum, presumably due to decoherence of the collective modes, there is good agreement between the measured and calculated microwave resonator spectrum.

The lowest frequency resonator considered in Refs. \cite{LiberskySM, Mckenzie, StampGSM} was $\sim 1$ GHz, which corresponds to the base temperature of the resonator ($T=50$ mK). As the mode softens to zero near the QCP, one must take into consideration other soft modes, such as soft photons and phonons, which couple to the order parameter. The effect of coupling an order parameter to other soft modes is often to drive a continuous transition first order \cite{Belitz}. Whether this happens in LiHoF$_4$, and the temperature and frequency at which these effects become important, is not clear to the author.

The experimental data for the magnetostatic modes calculated in this paper are not conclusive; however, they are consistent with the theoretical analysis. We leave refinements of the theory, and a more detailed experimental analysis of the modes, as a subject for future work.

\section{Conclusions}

We have presented a theory of quantum criticality and magnetostatic (Walker) modes in real quantum-Ising magnets with arbitrary shape. Our analysis of electronuclear-magnetostatic modes in terms of average demagnetization field and magnetization fluctuations is appropriate for microwave resonator experiments, in which the wavelength of the input and output photons is much larger than the sample size. Demagnetization-field fluctuations were found to gap the uniform (Kittel) mode in non-ellipsoidal samples; however, quantum criticality persists with the soft mode being a magnetostatic mode. Furthermore, we have analyzed a mode having lower energy then the soft mode that governs bulk quantum criticality. Experimental data was provided as evidence for these modes.

A magnetic material will divide itself into domains which null out surface effects, so that the system has a well defined thermodynamic limit, in accord with Griffiths' theorem \cite{GriffithsFE}. Note that Griffiths' theorem is valid in the absence of an applied field, and here we are considering the TFIM. In a system having strong uniaxial anisotropy, in which all but the longitudinal component of the dipolar interaction is quenched, Griffiths' theorem ought to be applicable in the absence of an applied longitudinal field. A primary new result of this paper is that quantum criticality in a non-ellipsoidal Ising magnet is governed by a magnetostatic soft mode, a result similar to Griffiths' theorem.

Both Griffiths' theorem, and this paper, indicate that shape effects may be neglected in efforts to understand quantum criticality. This justifies the use of a needle-shaped sample in calculations meant to capture bulk quantum-critical behavior. However, in an experiment, shape effects must be treated with care as they lead to additional modes not present in a needle. Note that the shape effects analyzed here are due to long-range dipole-dipole interactions; this is distinct from finite-size effects considered in renormalization group analyses of critical behavior. Domains and magnetostatic modes null out shape effects; however, quantum-critical behavior is still subject to finite-sized scaling, which gaps the soft mode and rounds off divergences in finite-sized systems \cite{GoldenfeldBook}.

The magnetostatic mode which softens to zero above and below $B_x^c$ provides a promising avenue for future research. We have shown that driven static magnetization fluctuations may cause this mode, and presented rudimentary experimental evidence for its existence. A more sophisticated theoretical analysis, and further experimental evidence, would be of interest.  

Quantum criticality is thought to be responsible for many of the exotic phases of matter and anomalous properties of systems at the forefront of condensed matter research. Quantum magnonics is evolving quickly, and associated quantum technologies are rapidly developing. In all of this, one must consider experimentally relevant details such as demagnetization fields and other shape effects. We have done so here by showing quantum criticality persists in non-ellipsoidal, dipole-dipole coupled, quantum-Ising magnet through the formation of a soft magnetostatic mode.

\section{Acknowledgements}

The author is indebted to M Libersky, DM Silevitch, and TF Rosenbaum for use of their experimental data. The author would also like to thank AA Geim and the Rosenbaum group for feedback and helpful discussions. Experimental work at the California Institute of Technology was supported by U.S. Department of Energy Basic Energy Sciences Grant No. DESC0014866.

\appendix

\section{Dynamic susceptibility of the quantum-Ising model}
\label{ap:TFIMsusc}

Consider the internal RPA susceptibility of a quantum-Ising material which follows from Eq. (\ref{eq:EOMInternal}). Setting $x=-i\omega/(\gamma \mu_0)$, the matrix on the left-hand side of Eq. (\ref{eq:EOMInternal}) is given by
\begin{align}
\widehat{A} = \left[ \begin{array}{ccc}
x  & H_{\emptyset}^z & 0 \\ 
-H_{\emptyset}^z & x & H^x - \Phi_{\emptyset} M^x  \\
0 & -H^x & x  \\ 
\end{array} \right],
\end{align}
and the RPA modes of the material follow from setting $\text{det}[A]=0$. One finds a ``longitudinal" zero mode, $\omega_{\parallel}=0$, where by ``longitudinal" we mean in the direction of the MF magnetization, and a transverse mode 
\begin{align}
\label{eq:omegaperp}
\omega_{\perp} = \pm (\gamma \mu_0)\sqrt{(H_{\emptyset}^z)^2 + H^x (H^x - \Phi_{\emptyset} M^x)}.
\end{align}
This collective mode softens to zero at the critical point of the TFIM.

Assuming $\omega \neq 0$, one may invert $A$ to obtain the internal susceptibility of the system
\begin{align}
\widehat{\chi}_{in}(\omega \neq 0) = \widehat{A}^{-1}\left[ \begin{array}{ccc}
0  & M^z & 0 \\ 
-M^z & 0 & M^x  \\
0 & -M^x & 0  \\ 
\end{array} \right].
\end{align}
One finds
\begin{align}
\widehat{\chi}_{in}(\omega \neq 0) = \left[ \begin{array}{ccc}
\chi^{xx}  & -i\kappa^{xy} & \chi^{xz} \\ 
i\kappa^{xy} & \chi^{yy} & -i \kappa^{yz}  \\
\chi^{xz} & i\kappa^{yz} &  \chi^{zz}  \\ 
\end{array} \right],
\end{align}
where we have suppressed the frequency dependence on the right-hand side. For the diagonal components of the susceptibility tensor we have
\begin{align}
\label{eq:chizz}
\chi^{xx} = \frac{(\gamma \mu_0)^2 M^zH_{\emptyset}^z}{\omega_{\perp}^2-\omega^2} \quad \text{and} \quad
\chi^{zz} = \frac{(\gamma \mu_0)^2 M^xH^x}{\omega_{\perp}^2-\omega^2},
\end{align}
and
\begin{align}
\chi^{yy} 
=\frac{M^x}{H^x} \frac{\omega_{\perp}^2}{\omega_{\perp}^2-\omega^2}
=\frac{M^z}{H_{\emptyset}^z} \frac{\omega_{\perp}^2}{\omega_{\perp}^2-\omega^2}.
\end{align}
For the off-diagonal components of the susceptibility one finds
\begin{align}
\chi^{xz} = \chi^{zx} = -\frac{(\gamma \mu_0)^2 H_{\emptyset}^zM^x}{\omega_{\perp}^2-\omega^2}.
\end{align}
The remaining off-diagonal components are $\chi^{xy} = -\chi^{yx} = -i \kappa^{xy}$ and $\chi^{yz} = -\chi^{zy} = -i \kappa^{yz}$, with
\begin{align}
\kappa^{xy} = \frac{\omega \gamma \mu_0 M^z}{\omega_{\perp}^2-\omega^2} \quad \text{and} \quad
\kappa^{yz} = \frac{\omega \gamma \mu_0 M^x}{\omega_{\perp}^2-\omega^2}.
\end{align}

As one approaches the critical point of the TFIM, $M^z \rightarrow 0$ and $\omega_{\perp} \rightarrow 0$, hence both $\chi^{xx}$ and $\chi^{yy}$ vanish. In a spectroscopy experiment, one must measure $\chi^{zz}$ to see resonant absorption by the soft mode \cite{MckenzieStamp, LiberskySM}. The antisymmetry of $\chi^{xy}$ and $\chi^{yz}$ lead to a cancellation in the Walker-mode equation given by Eq. (\ref{eq:Walker}). As $\chi^{xz} = \chi^{zx}$, the Walker-mode equation for the TFIM contains a term not present in the usual Walker-mode equation in which the applied ac field is orthogonal to the direction of magnetization \cite{Walker1, StancilPrabhakarBook}.

We have derived the internal RPA susceptibility of the TFIM. The MF result, which follows from Eq. (\ref{eq:EOMMF}) is similar; one must simply replace the transverse RPA mode with the MF result, $\omega_{\perp} \rightarrow \omega_{\perp}^{MF} = \gamma \mu_0 \sqrt{(H^x)^2+(H_{\emptyset}^z)^2}$. In the absence of the transverse field, considering an ac field orthogonal to the direction of magnetization, one obtains the usual Polder susceptibility tensor ($\chi_{MF}=\chi_{MF}^{xx}=\chi_{MF}^{yy}$ and $\kappa_{MF}=\kappa_{MF}^{xy}$)
\begin{align}
\widehat{\chi}_{MF}(\omega \neq 0) = \left[ \begin{array}{cc}
\chi_{MF}  & -i\kappa_{MF}  \\ 
i\kappa_{MF} & \chi_{MF}   \\ 
\end{array} \right],
\end{align}
where 
\begin{align}
\chi_{MF} = \frac{\omega_0\omega_M}{\omega_0^2-\omega^2} \quad \text{and} \quad
\kappa_{MF} = \frac{\omega \omega_M}{\omega_0^2-\omega^2},
\end{align}
with $\omega_0 = \gamma \mu_0 H_{\emptyset}^z$ and $\omega_M = \gamma \mu_0 M^z$.

We have determined the dynamic susceptibility of the quantum-Ising model making use of the EOM of the spin operators. Alternately, one may consider the connected, imaginary time ordered, correlation function, $G_{ij}^{\mu \nu}(\tau) = -\langle T_{\tau} \delta J_i^{\mu} (\tau) \delta J_j^{\nu}(0) \rangle$. Transforming to momentum and Matsubara frequency space, this function is related to the dynamic spin susceptibility by $\chi_{\boldsymbol{k},J}^{\mu \nu}(\omega) = -\beta G_{\boldsymbol{k}}^{\mu \nu}(i\omega_r \rightarrow \omega + i0^+)$. The dimensionless susceptibility of the material is then $\chi_{\boldsymbol{k}}^{\mu \nu}(\omega) = 4\pi J_D \rho_s \chi_{\boldsymbol{k},J}^{\mu \nu}(\omega)$ In the RPA, at zero wavevector, one finds the longitudinal component of the susceptibility to be 
\begin{align}
\label{eq:xRPA}
\chi_{RPA}^{zz}(\omega \neq 0) 
= \frac{\chi_{\emptyset}^{zz}(\omega \neq 0)}{1-\Phi_{\emptyset}\chi_{\emptyset}^{zz}(\omega \neq 0)},
\end{align}
where $\chi_{\emptyset}^{zz}$ is the MF susceptibility of a needle, and $\Phi_{\emptyset} = V_{\emptyset}/(4\pi J_D \rho_s)$. The poles of Eq. (\ref{eq:RPA}) determine the RPA modes of the material, and their residues determine spectral weights.

Indeed, in terms of MF eigenstates and matrix elements,
\begin{align}
\mathcal{H}_{MF,i} &= \sum_{n} E_n | X_{n,i} \rangle \langle X_{n,i}| \quad \text{and} \quad \\ \nonumber
c_{mn} &= \langle X_{m,i} | J_i^z | X_{n,i} \rangle
\end{align}
with $\{X_{n,i}\}$ being single-ion Hubbard operators, one finds
\begin{align}
\chi_{\emptyset}^{zz}(\omega \neq 0) = 
\sum_{n > m} |c_{mn}|^2 p_{mn} \frac{4\pi J_D \rho_s 2 E_{nm}}{E_{nm}^2-(\omega + i0^+)^2},
\end{align}
where the $p_{mn} = p_m - p_n$ are differences between population factors $p_n = e^{-\beta E_n}/Z_{MF}$. A straightforward calculation shows the results for the susceptibility, and the RPA mode energy, are equivalent to the results given in Eqs. (\ref{eq:omegaperp}) and (\ref{eq:chizz}). The Green function approach is advantageous when dealing with electronuclear systems as it avoids the RPA decoupling of the electronic and nuclear spins inherent in the EOM approach.

\section{Electronuclear dynamics and frequency pulling}
\label{ap:FP}

Early work on electronuclear dynamics was carried out be de Gennes \textit{et al} \cite{deGennesNMR}. They found that interactions between nuclear moments mediated by the electronic spins (the Suhl-Nakamura interaction \cite{Suhl, Nakamura}) lead to a reduction in the energy of the nuclear mode, known as frequency pulling. Their results were used by Blocker to analyze magnetostatic modes in electronuclear systems \cite{Blocker}. These authors consider a ferromagnet magnetized in the $z$ direction by an applied field, with an isotropic hyperfine interaction, subject to an ac field orthogonal to the direction of magnetization. For reference, we review the frequency pulling result making use of the notation in this paper.

Consider the dipolar Ising model in a longitudinal field with an isotropic hyperfine interaction
\begin{align}
\mathcal{H} = -\frac{1}{2} &\sum_{i \neq j} V_{ij}^{zz} J_i^z J_j^z \\ \nonumber
&+ h\sum_i (J_i^z - \gamma I_i^z)
+ A \sum_i \boldsymbol{J}_i \cdot \boldsymbol{I}_i,
\end{align}
where $\gamma = |\gamma_n/\gamma_e|$ is the ratio of nuclear and electronic gyromagnetic rations, and the interaction between spins is dipolar, $V_{ij}^{zz} = J_D D_{ij}^{zz}$. The total moment at each site is given by $\boldsymbol{\mu} = \gamma_e \boldsymbol{J} + \gamma_n \boldsymbol{I}$ ($\hbar=1$), and the local magnetization is $\boldsymbol{M} = \rho_s \boldsymbol{\mu} = \boldsymbol{M}_e + \boldsymbol{M}_n$. Assuming a uniformly magnetized ellipsoid and treating the interaction in MF theory, the single-ion Hamiltonian may be written
\begin{align}
\mathcal{H}_{MF} = -\frac{\mu_0}{\rho_s} 
[ H_{loc}^z M_e^z + H_n^z M_n^z + \alpha \boldsymbol{M}_e \cdot \boldsymbol{M}_n] .
\end{align}
The local field experienced by the electronic moments is
\begin{align}
H_{loc}^z = -\frac{h}{\mu_0 \gamma_e} + \Phi_0 M_e^z,
\end{align}
where $\Phi_0 = D_0^{zz}/4\pi \rho_s$ (see Eq. \ref{eq:Hdip2}). The longitudinal nuclear field and hyperfine coupling are
\begin{align}
H_n^z = -\frac{h}{\gamma_e \mu_0}
\quad \text{and} \quad
\alpha = -\frac{A}{ \mu_0 \rho_s \gamma_e \gamma_n}.
\end{align}
Note that $\gamma_e = -g \mu_B < 0$, so that $\alpha$ and the fields are greater than zero. We proceed by analyzing the EOM for the coupled electronic and nuclear magnetizations.

The coupled equations for the electronic and nuclear magnetizations are
\begin{align}
\frac{d}{dt} \boldsymbol{M}_e(t) &= 
\mu_0 \gamma_e \boldsymbol{M}_e(t) \times [H_{loc}^z + \alpha \boldsymbol{M}_n(t)] \quad \text{and}
\\ \nonumber
\frac{d}{dt} \boldsymbol{M}_n(t) &= 
\mu_0 \gamma_n \boldsymbol{M}_n(t) \times [H_n^z + \alpha \boldsymbol{M}_e(t)].
\end{align}
In a transverse ac field, we assume small transverse motions of the electronic spins so that $\boldsymbol{M}_e(t) = \boldsymbol{M}_e +\delta \boldsymbol{m}_e (t) = (\delta m_e^x(t),\delta m_e^y(t),M_e^z)$. At the MF level of approximation, we neglect fluctuations in the local field $H_{loc}^z$ (see Eq. \ref{eq:EOMMF}). Writing the nuclear spin magnetization as $\boldsymbol{M}_n(t) = \boldsymbol{M}_n + \delta \boldsymbol{m}_n(t)$, the linearized EOM for the electronic magnetization is
\begin{align}
\label{eq:me}
&\boldsymbol{M}_e \times (H_{loc}^z + \alpha \boldsymbol{M}_n) = 0
\quad \text{and} \\ \nonumber
&\frac{d}{dt} \delta \boldsymbol{m}_e = \mu_0 \gamma_e [\delta \boldsymbol{m}_e 
\times (H_{loc}^z + \alpha \boldsymbol{M}_n) + \alpha \boldsymbol{M}_e 
\times  \delta \boldsymbol{m}_n],
\end{align}
where in the the final line we have suppressed the time dependence of the electronic and nuclear magnetization fluctuations, $\delta \boldsymbol{m}_{e,n}(t)$. 

As $\boldsymbol{M}_e = (0,0,M_e^z)$, the first expression in Eq. (\ref{eq:me}) tells us that $\boldsymbol{M}_n = (0,0,M_n^z)$. In a transverse ac field, we consider transverse fluctuations about the average magnetization, $\boldsymbol{M}_n(t) = (\delta m_n^x(t),\delta m_n^y(t),M_n^z)$. It is assumed by de Gennes \textit{et al.} that the electrons follow the field $H_{loc}^z + \alpha \boldsymbol{M}_n(t)$ adiabatically so that $\boldsymbol{M}_e(t) \parallel [H_{loc}^z + \alpha \boldsymbol{M}_n(t)]$ \cite{deGennesNMR}. The equation for the electronic fluctuations is then
\begin{align}
\delta \boldsymbol{m}_e \times (H_{loc}^z + \alpha \boldsymbol{M}_n)
 + \alpha \boldsymbol{M}_e \times  \delta \boldsymbol{m}_n = 0.
\end{align}
In terms of the circularly polarized wave $\delta m_e^+ = \delta m_x + i \delta m^y$, one finds
\begin{align}
\label{eq:fluc}
\delta m_e^+ = \frac{\alpha M_e^z}{H_{loc}^z+\alpha M_n^z} \delta m_n^+.
\end{align}
This is equivalent to Eq. 2.2 of de Gennes \textit{et al.} with two caveats: 1) The local field considered by de Gennes \textit{et al.} includes an applied field and an anisotropy field. Here, we consider an Ising system, and the local field includes an applied field as well as the dipolar MF. 2) The hyperfine correction to the local field in the denominator of Eq. (\ref{eq:fluc}) is not included in the work of de Gennes \textit{et al.}. If the hyperfine interaction is weak, this term will be negligibly small and the two equations are equivalent; however, in materials with strong hyperfine couplings, such as LiHoF$_4$, this term should not be neglected. Making use of Eq. (\ref{eq:fluc}), we proceed by analyzing the EOM for the nuclear magnetization.

The linearized EOM for fluctuations in the nuclear magnetization is
\begin{align}
\frac{d}{dt} \delta \boldsymbol{m}_n = \mu_0 \gamma_n [\delta \boldsymbol{m}_n 
\times (H_n^z + \alpha \boldsymbol{M}_e) +  \alpha \boldsymbol{M}_n 
\times  \delta \boldsymbol{m}_e]
\end{align}
For a circularly polarized nuclear magnetic fluctuation, $\delta m_n^+=\delta m_n^x+i \delta m_n^y$, one finds
\begin{align}
\frac{d}{dt} \delta m_n^+ = i\mu_0\gamma_n 
[\delta m_e^+ \alpha M_n^z  - \delta m_n^+ (H_n^z + \alpha M_e^z)].
\end{align}
Considering a single Fourier component of the nuclear magnetization fluctuation, $\delta m_n^+(t) = \delta m_n^+ e^{i\omega t}$, and making use of equation (\ref{eq:fluc}), we find the nuclear resonance frequency to be
\begin{align}
\omega_n &= -\mu_0 \gamma _n  \biggr[
H_n^z+ \alpha M_e^z \biggr(1-\eta\frac{M_n^z}{M_e^z}\biggr)\biggr] 
 \\ \nonumber &\quad \text{with} \quad
\eta = \frac{\alpha M_e^z}{H_{loc}^z + \alpha M_n^z}.
\end{align}
This is in agreement with Eq. 2.4 of de Gennes \textit{et al.} apart from the additional hyperfine contribution in the denominator of $\eta$. As previously noted, if the hyperfine coupling is small $\alpha M_n^z$ may be neglected. In the absence of the interactions between the nuclear spins mediated by the electronic spins (the Suhl-Nakamura interaction), the nuclear spins would precess about the applied field and electronic MF with frequency $\omega_n = -\mu_0 \gamma_n (H_n^z+\alpha M_e^z)$. The Suhl-Nakamura interaction leads to a reduction in the frequency of the nuclear mode known as frequency pulling. 

In terms of the parameters of the original spin Hamiltonian, the nuclear resonance frequency is (note $\langle J^z \rangle_{MF} \le 0$ and $\langle I^z \rangle_{MF} \ge 0$) 
\begin{align}
\omega_n = -\gamma h + A \langle J^z \rangle_{MF} 
\biggr[\frac{h_{MF}}{h_{MF}+ A \langle I^z \rangle_{MF}}\biggr],
\end{align}
where $h_{MF} = h-V_0^{zz} \langle J^z \rangle_{MF}$.
One may make use of the adiabatic approximation to analyze electronuclear modes in the TFIM with an anisotropic hyperfine interaction; however, the resulting equations become rather complicated. In this paper, we have considered the $6 \times 6$ EOM for the coupled electronic and nuclear spins, subject to a longitudinal ac field, and solved for the electronuclear modes numerically.

\bibliography{MM.bib}

\end{document}